\documentclass[journal,draft,onecolumn,12pt,twoside]{IEEEtran}
	\usepackage{cite}

	\usepackage[cmex10]{amsmath}

	\usepackage{array}
	\usepackage{mdwmath}
	\usepackage{mdwtab}
	\usepackage{mathtools}

	\usepackage{url}
	
	\usepackage{graphics} 
	\usepackage{epsfig} 
	\usepackage{amssymb}  
	\usepackage{hyperref} 
	\usepackage{mathtools}
	\usepackage{subfig}
	\usepackage{verbatim}
	\usepackage{color}
	\usepackage{multirow} 

	\newtheorem{theorem}{Theorem}
	\newtheorem{definition}{Definition}
	\newtheorem{lemma}{Lemma}

		\newtheorem{remark}{Remark}

\newcommand{\eq}[1]{\begin{align}#1\end{align}}
\newcommand{\seq}[1]{\begin{subequations}#1\end{subequations}}

\newcommand{\E}{\mathbb{E}}
\newcommand{\cX}{\mathcal{X}}
\newcommand{\cA}{\mathcal{A}}

\newcommand{\cP}{\Delta}

\newcommand{\cH}{\mathcal{H}}

\newcommand{\ha}{\hat{a}}

\newcommand{\cN}{\mathcal{N}}

\newcommand{\defeq}{\buildrel\triangle\over =}

\newcommand{\pushright}[1]{\ifmeasuring@ #1 \else\omit\hfill$\displaystyle#1$\fi\ignorespaces}
\newcommand{\pushleft}[1]{\ifmeasuring@ #1 \else\omit$\displaystyle#1$\hfill\fi\ignorespaces}
\newcommand{\nn}{\nonumber}

\begin{document}
	%
	\title{Sequential decomposition of repeated games with asymmetric information and dependent states}
	%
	%
	%
	\author{Deepanshu~Vasal%
	\thanks{Deepanshu Vasal is with the Department
	of Electrical and Computer Engineering, and with Simons Center for Mathematics of Networks, University of Texas, Austin, TX, USA. Email: dvasal at utexas.edu.} 
	\thanks{This research was funded by Department of Defense $\#$W911NF1510225 to The University of Texas
at Austin and Simon's Foundation grant 26-7523-99.}%
	}
	
	
	%
	\maketitle
	\abstract
We consider a finite horizon repeated game with $N$ selfish players who observe their types privately and take actions, which are publicly observed. Their actions and types jointly determine their instantaneous rewards. In each period, players jointly observe actions of each other with delay 1, and private observations of the state of the system, and get an instantaneous reward which is a function of the state and everyone's actions. The players' types are static and are potentially correlated among players. 
 An appropriate notion of equilibrium for such games is Perfect Bayesian Equilibrium (PBE) which consists of a strategy and a belief profile of the players which is coupled across time and as a result, the complexity of finding such equilibria grows double-exponentially in time. We present a sequential decomposition methodology to compute \emph{structured perfect Bayesian equilibria} (SPBE) of this game, introduced in~\cite{VaAn15arxiv}, where equilibrium policy of a player is a function of a common belief and a private state. This methodology computes SPBE in linear time. In general, the SPBE of the game problem exhibit \textit{signaling} behavior, i.e. players' actions reveal part of their private information that is payoff relevant to other players.

		\section{Introduction}

\normalfont{Information} asymmetry among strategic agents is an important topic, which has seen some very influential works such as~\cite{Ak70} and~\cite{Sp73}. Akerlof in~\cite{Ak70} and Spence in~\cite{Sp73} modeled a market of cars and a job market, respectively, as instances of information asymmetry in a game, and show interesting behavior of strategic agents derived from these models. Specifically, Akerlof in~\cite{Ak70} showed that in a market of cars, where the quality of car is known only to the seller, lower prices can drive out good cars from the market. Spence in~\cite{Sp73} showed that in equilibrium in a job market, a candidate can `signal' her higher productivity to a potential employer by opting for higher education credentials. While these works showed very interesting and relevant phenomena for \emph{static} information asymmetry, in the real world however, there exists many such, and even more complicated decision making scenarios which involves strategic decision makers with \emph{dynamically evolving} information asymmetry. Some instances of such systems include: (a) in cyber-physical systems, many cyber and physical devices are connected to each other which have different information and they make a decision to optimize their performance objectives; (b) in a wind energy market a wind energy producer observes its own wind production privately and publicly observes the output of other producers which also determine the prices, and its objective is to generate output that maximizes its revenue; (c) in a social network, people have private opinions about a topic and also publicly observe actions of others, based on which they make a decision to maximize their utility. All such scenarios can be modeled as a dynamic game of asymmetric information\footnote{Sometimes also referred to as dynamic games of incomplete/imperfect information.}~\cite{FuTi91}. Such problems are gaining more interest with applications such as alpha-go by Deepmind~\cite{Sietal17} for the symmetric information game Go, and for asymmetric/imperfect information games such as Texas Hold'em in~\cite{BrSa17}.

Dynamical systems with strategic players are modeled as dynamic stochastic games, introduced by Shapley in~\cite{Sh53}. Discrete-time dynamic games with Markovian structure have been studied extensively to model many practical applications, in engineering as well as economics literature~\cite{BaOl98, FiVr12}. In dynamic games with perfect and symmetric information, subgame perfect equilibrium (SPE) is an appropriate equilibrium concept and there exists a backward recursive methodology to find all the SPEs of these games (see~\cite{OsRu94, FuTi91book, samuelson2006} for a more elaborate discussion). Maskin and Tirole in \cite{MaTi01} introduced the concept of Markov perfect equilibrium (MPE) for dynamic games with symmetric information, where equilibrium strategies are dependent on some payoff relevant Markovian state of the system, rather than on the entire history. 
In dynamic games of asymmetric information, there are strategic players who are affected by an underlying process that is dynamically evolving, and the players make asymmetric observations about that process. In such games and more generally in any dynamic multi-agent decision problem with asymmetric information, a player's action not just either \emph{explore} or \emph{exploit} the system\footnote{Exploitation refers to making a decision based on   whereas exploration refers to taking action that improves the current estimate of the state of the system even at some cost in the present, but that improves future reward.}, as it happens in a single agent problem, but also \emph{signal} i.e. reveal part of its private information to the other players that is payoff relevant to them.\footnote{Equivalently, signaling occurs in such decision problems if players' beliefs on a payoff relevant state are strategy dependent i.e. they depend on the strategies of one or more players.}
Some appropriate notions of equilibrium for such games is Perfect Bayesian Equilibrium (PBE) or Sequential Equilibrium (SE)~\cite{OsRu94,FuTi91} which involve an equilibrium strategy profile and an equilibrium belief profile of all the players, among other refinements. In these equilibria, the equilibrium strategies and beliefs are coupled together through a joint fixed-point equation in the space of strategies and beliefs for all players and for all histories of the game.\footnote{In comparison, Nash equilibrium for a static game is a fixed-point equation in the space of probability measures on possible actions of the players~\cite{OsRu94}} Since the history of such games grows exponentially, the complexity of finding equilibria of such game grows double exponentially in time, rendering such problems intractable. We refer the reader to the Introduction section of~\cite{VaSiAn16arxiv} for a thorough introduction and a brief literature survey on dynamic games with asymmetric information.

	Recently, there have been a number of results on finding an information state for different classes of such games that decomposes these games across time (in an analogous way a dynamic program decomposes a dynamic optimization problem), and thus reduces the complexity of finding these equilibria from double-exponential to linear in time. Authors in~\cite{VaAn16,VaSiAn16arxiv}, and independently, authors in~\cite{OuTaTe17, Ta17}, presented such a sequential decomposition for games to find \emph{structured perfect Bayesian equilibrium (SPBE)} and \emph{common information based perfect Bayesian equilibrium (CIB-PBE)}, respectively, where each player has a type or a state that evolves (conditionally) independently of other players' types in a Markovian way. Authors in~\cite{VaAn16cdc} extended those results to LQG games (i.e. with linear state update, quadratic instantaneous costs and Gaussian random variables), and to games with conditionally independent hidden Markovian types in~\cite{VaAn16allerton}, where instead of perfectly observing its own type, each player makes independent noisy observations about it, respectively. Authors in~\cite{TaOuTe16cdc} considered such dynamic game with a system state and {delayed information sharing pattern} where each players learns every agents private observations and actions with delay of $d-$steps. Authors in~\cite[ch. 5]{Ta17} generalized that dynamic game with a Markovian state where players make a common and private observations of the state of the system, where these observations are conditionally independent among players, conditioned on the current state and previous action. In this paper, we consider a model that adds to this literature, where there exists players' types that are static and \emph{arbitrary correlated} among the players.  

In this paper, similar to~\cite{VaSiAn16arxiv}, we present a backward-forward methodology to compute SPBE of the game, where each player's strategy is a function of the same common information belief state and private state. These equilibria of the game are analogous to Markov Perfect equilibria (MPE)~\cite{MaTi01} of symmetric information games.

The paper is structured as follows. In Section~\ref{sec:Model}, we present the general model. In Section~\ref{sec:Motivation}, we define information belief states, which act as the motivation for the structure of the equilibrium policies. In Section~\ref{sec:Game_finite}, present the backward-forward methodology to compute SPBE of the game in linear time. We discuss some remarks in Section~V. 
We conclude in section~\ref{sec:Conclusion}.

\subsection{Notation}
We use uppercase letters for random variables and lowercase for their realizations. For any variable, subscripts represent time indices and superscripts represent player identities. We use notation $ -i$ to represent all players other than player $i$ i.e. $ -i = \{1,2, \ldots i-1, i+1, \ldots, N \}$. We use notation $a_{t:t'}$ to represent vector $(a_t, a_{t+1}, \ldots a_{t'})$ when $t'\geq t$ or an empty vector if $t'< t$. We use $a_t^{-i}$ to mean $(a^1_t, a^2_t, \ldots, a_t^{i-1}, a_t^{i+1} \ldots, a^N_t)$ . We remove superscripts or subscripts if we want to represent the whole vector, for example $ a_t$  represents $(a_t^1, \ldots, a_t^N) $. In a similar vein, for any collection of finite sets $(\cX^i)_{i \in \mathcal{N}}$, we denote $\times_{i=1}^N \cX^i$ by $\cX$. We denote the indicator function of any set $A$ by $I_{A}(\cdot)$. 
For any finite set $\mathcal{S}$, $\Delta(\mathcal{S})$ represents space of probability measures on $\mathcal{S}$ and $|\mathcal{S}|$ represents its cardinality. We denote by $P^g$ (or $E^g$) the probability measure generated by (or expectation with respect to) strategy profile $g$. We denote the set of real numbers by $\mathbb{R}$. For a probabilistic strategy profile of players $(\beta_t^i)_{i\in \mathcal{N}}$ where probability of action $a_t^i$ conditioned on $(a_{1:t-1},x^i)$ is given by $\beta_t^i(a_t^i|a_{1:t-1},x^i)$, we use the short hand notation $\beta_t^{-i}(a_t^{-i}|a_{1:t-1},x^{-i})$ to represent $\prod_{j\neq i} \beta_t^j(a_t^j|a_{1:t-1},x^j)$.   
All equalities and inequalities involving random variables are to be interpreted in \emph{a.s.} sense.

\section{General model}
\label{sec:Model}
We consider a discrete-time dynamical system with $N$ selfish players in the set $ \cN \defeq \{1,2, \ldots N \}$. {We consider finite horizon $\mathcal{T} \defeq \{1, 2, \ldots T\}$ with perfect recall}. The system state is $x = (x^1,\ldots,x^N)$, where $x \in \mathcal{X}$ is distributed as
\eq{
P(x) &= Q^x(x)
}
where $a_t = (a_t^1, \ldots, a_t^N)$, and $a_t^i$ is the action taken by player $i$ at time $t$. At start of period $t$, players jointly observe $a_{t-1}$, and make private observations $x$, where $x^i \in \cX^i$ is the private observation of player $i$.$\forall t \in {1, \ldots T} $,

Player $i$ takes action $a_t^i \in \cA^i$ at time $t$ upon observing $a_{1:t-1}$ which is common information among players, and $x^i$ which is player $i$'s private information. The sets $\cA^i, \cX^i$ are assumed to be finite. Let $g^i = (g^i_t)_t$ be a probabilistic strategy of player $i$ where $g^i_t : \cA^{t-1}\times (\cX^{i}) \to \cP(\cA^i)$ such that player $i$ plays action $a_t^i$ according to $A_t^i \sim g^i_t(\cdot|a_{1:t-1}, x^i)$. Let $g:= (g^i)_{i\in \mathcal{N}}$ be a strategy profile of all players. At the end of interval $t$, player $i$ gets an instantaneous reward $R_t^i(X,A_t)$. The objective of player $i$ is to maximize its total expected reward in a \emph{perfect} sense
\eq{
 J^{i,g} := \E^g \left[ \sum_{t=1}^T R_t^i(X,A_t) \right] .
}
 
 \section{Structural Results}
 \label{sec:Motivation}
 The problem described in previous section is a repeated game with asymmetric information where a player's strategy is of the form $a_t^i \sim g_t^i(\cdot|a_{1:t-1},x^i)$. Equivalent of such decentralized problems when players are cooperative and have the same objective can not be solved using classical tools from the theory of Markov Decision Processes (MDP)~\cite{KuVa86}. However, there exists some key ideas in the literature such as agent-by-agent approach~\cite{Ho80} and common information approach~\cite{NaMaTe13} that present structural results of the optimum policies for a class of such systems with non-classical information structure and cooperative players. Such techniques have been used in dynamic team problems such as~\cite{Ma13, MaTe08,VaAn14}. For the problem described in previous section, we use an approach inspired by~\cite{NaMaTe13}.

For every agent $i\in\cN$ and any policy profile $g$, let $\pi_t\in \cP(\cX)$ be a common belief on $x$ conditioned on the common information $a_{1:t-1}$, defined as follows.
\eq{
\pi_t(x) &:=P^{g}(x=x|a_{1:t-1}) 
}

\subsection{Common agent approach}
An alternative way to view the problem is as follows. As is done in common information approach~\cite{NaMaTe13}, at time $t$, a fictitious common agent observes the common information $a_{1:t-1}$ and generates prescription functions $\gamma_t = (\gamma_t^i)_{i\in\cN} = \psi_t[a_{1:t-1}]$. Player $i$ uses these prescription functions $\gamma_t^i$ to operate on its private information $x^i$ to produce its action $a_t^i$, i.e. $\gamma_t^i : x^i \to \cP(\cA^i)$ and $a_t^i \sim\gamma_t^i(\cdot|x^i) = \psi[a_{1:t-1}](\cdot|x^i)$. It is easy to see that for any $g$ policy profile of the players there exists an equivalent $\psi$ profile of the common agent (and vice versa) that generates the same control actions for every realization of the information of the players.

We define a special type of common agent's policy as follows. We call a common agent's policy be of type $\theta$ if the common agent observes the common belief $\pi_t$ derived from the common observation $a_{1:t-1}$, and generates prescription functions $\gamma_t = (\gamma_t^i)_{i\in\cN} = \theta_t[\pi_t]$. Player $i$ uses these prescription function $\gamma_t^i$ to operates on its private type $x^i$ to produce its action $a_t^i$, i.e. $\gamma_t^i : \cP(\cX)\to \cP(\cA^i)$ and $a_t^i \sim\gamma_t^i(\cdot|x^i) = \theta[\pi_t](\cdot|x^i)$. 
Equivalently, we call a common agent's policy be of type $\theta$ if for all $i\in\cN$ and for all time $t$, player $i$'s action $a_t^i$ depends on its information $(a_{1:t-1}, x^i)$ through the belief states $\pi_t(x) = P^{\theta}(x=x|a_{1:t-1})$ and $x^i$.

In the next lemma we show that for any given $\theta$ policy, the belief states $\pi_t$ can be updated recursively as follows. Let  $\pi_1(x) := Q^x(x)$.

\begin{lemma} For any given policy of type $\theta$, there exists update functions $F_t$, independent of $\theta$, such that
	\eq{
	\pi_{t+1} = F_t(\pi_t,\gamma_t,a_{t}). \label{eq:piiupdate_a}
	}
	\label{lemma:C1}
	\end{lemma}
	\begin{IEEEproof}
	Please see Appendix~\ref{app:lemma_update}.
	\end{IEEEproof}
	
\section{Repeated games with asymmetric information} 
\label{sec:Game_finite}
\subsection{Solution concept: PBE}
We introduce perfect Bayesian equilibrium (PBE) as an appropriate equilibrium concept for the game considered.
Any history of this game at which players take action is of the form $h_t = (x,a_{1:t-1})$. Let $\cH_t$ be the set of such histories at time $t$. At any time $t$ player $i$ observes $h^i_t = (x^{i},a_{1:t-1})$ and all players together observe $h^c_t = (a_{1:t-1})$ as common history. Let $\cH^i_t$ be the set of observed histories of player $i$ at time $t$ and $\cH^c_t$ be the set of common histories at time $t$. An appropriate concept of equilibrium for such games is PBE \cite{FuTi91book} which consists of a pair $(\beta^*,\mu^*)$ of strategy profile $\beta^* = (\beta_t^{*,i})_{t \in \mathcal{T},i\in\cN}$ where $\beta_t^{*,i} : \cH_t^i \to \cP(\cA^i)$ and a belief profile $\mu^* = (^i\mu_t^{*})_{t \in \mathcal{T},i\in\cN}$ where $^i\mu_t^{*}: \cH^i_t \to \cP(\cH_t)$ that satisfy sequential rationality so that $\forall i\in\cN, t \in \mathcal{T},  h^{i}_t \in \cH^i_t, {\beta^{i}}$
\eq{
&\E^{(\beta^{*,i} \beta^{*,-i},\, \mu^*)}\left\{ \sum_{n=t}^T R_n^i(X, A_n)\big\lvert  h^i_t\right\} \geq \E^{({\beta}^{i} \beta^{*,-i},\, \mu^*)}\left\{ \sum_{n=t}^T R_n^i(X, A_n)\big\lvert  h^i_t\right\}, \;\; \;\;   \label{eq:seqeq}
} 
and the beliefs are updated using Bayes' rule, whenever possible. 

In this paper, we first define a common belief $\pi_t^*$ as a belief on $x$. Then player $i$ derives its equilibrium belief $^i\mu_t^*$ on $(x^{-i})$ by conditioning $\pi_t^*$ on its private type $x^i$ i.e. $^i\mu_t^*(x^{-i}) = \pi_t^*(x^{-i}|x^i)$. 
We define structured perfect Bayesian equilibrium as follows.
\begin{definition}[SPBE]
	A structured perfect Bayesian equilibrium (SPBE) is a PBE of the considered dynamic game where at any time $ t $, for any agent $ i $, its equilibrium strategy $ \beta_t^{\ast,i} $ depends on player i's information $(a_{1:t-1},x^i)$ through the common belief and its private type $x^i$.
\end{definition}
We conjecture that as in~\cite{VaSiAn16arxiv}, such \emph{structured} strategies form a rich class where any expected reward profile of the players that can be generated from any general strategy profile can also be generated using such structured strategy profile.

In the following we present a backward-forward methodology to compute SPBE of this game.

\subsubsection{Backward Recursion}
In this section, we define an equilibrium generating function $\theta=(\theta^i_t)_{i\in\cN,t\in\mathcal{T}}$ and a sequence of functions 
$(V_t^i)_{i\in \cN, t\in \{ 1,2, \ldots T+1\}}$, where $V_t^i : \cP(\cX ) \times \cX^i \to \mathbb{R}$, in a backward recursive way, as follows.
\begin{itemize}
\item[1.] Initialize $\forall \pi_{T+1}\in \cP(\cX), x^i\in \cX^i$,
\eq{
V^i_{T+1}(\pi_{T+1},x^i) := 0.   \label{eq:VT+1}
}
\item[2.] For $t = T,T-1, \ldots 1, \ \forall \pi_t \in \cP(\cX )$, let $\theta_t[\pi_t] $ be generated as follows. Set $\tilde{\gamma}_t = \theta_t[\pi_t]$, where $\tilde{\gamma}_t$ is the solution, if it exists\footnote{Existence of general solution of this per stage fixed-point equation is discussed in Remark~\ref{remark:existence}.}, of the following fixed-point equation, $\forall i \in \cN,x^i\in \cX^i$,
  \eq{
 \tilde{\gamma}^{i}_t(\cdot|x^i) \in &\arg\max_{\gamma^i_t(\cdot|x^i)} \E^{\gamma^i_t(\cdot|x^i) \tilde{\gamma}^{-i}_t,\,\pi_t} \left\{ R_t^i(X,A_t) +    V_{t+1}^i (F_t(\pi_t, \tilde{\gamma}_t,A_t),x^i) \big\lvert x^i \right\} , \label{eq:m_FP}
  }
 where expectation in (\ref{eq:m_FP}) is with respect to random variables $(X,A_t)$ through the measure\\
$\sum_{x^{-i},x}\pi_t(x^{-i}|x^i)\gamma^i_t(a^i_t|x^i) \tilde{\gamma}^{-i}_t(a^{-i}_t|x^{-i})$, $F_t$ is defined in Lemma~\ref{lemma:C1}. Furthermore, set
  \eq{
  &V^i_t(\pi_t,x^i) :=\E^{\tilde{\gamma}^{i}_t(\cdot|x^i) \tilde{\gamma}^{-i}_t,\, \pi_t}\left\{ {R_t}^i (x,A_t)+   V_{t+1}^i (F_t(\pi_t, \tilde{\gamma}_t,A_t), x^i)\big\lvert  x^i \right\}.  \label{eq:Vdef}
   }
   \end{itemize}
It should be noted that (\ref{eq:m_FP}) is a fixed-point equation where the solution of the above fixed-point equation $\tilde{\gamma}^i_t$ appears in both, the left-hand-side and the right-hand-side of the equation. However, it is not the outcome of the maximization operation as in a best response equation similar to that of a Bayesian Nash equilibrium.

\subsubsection{Forward Recursion}
Based on $\theta$ defined above in (\ref{eq:VT+1})--(\ref{eq:Vdef}), we now construct a set of strategies $\beta^*$ and beliefs $\mu^*$ for the game $\mathfrak{D}$ in a forward recursive way, as follows, where $\beta_t^{i,*}:\mathcal{A}^{t-1}\times\mathcal{X}^{i}\to\Delta(\mathcal{A}^i)$ and $^i\mu_t^*:\mathcal{A}^{t-1}\times\mathcal{X}^{i}\to\Delta(\mathcal{X})$. We first define the update function of equilibrium common belief $\pi_t^*: \mathcal{A}^{t-1}\to\Delta(\mathcal{X})$, where for every private history $(a_{1:t-1},x^i)$ of player $i$, it's equilibrium belief $^i\mu_t^*$ is constructed by conditioning common belief $\pi_t^*$ on its private state $x^{i}$. 
\begin{itemize}
\item[1.] Initialize at time $t=0, \forall i\in\cN$,
\eq{
\pi_1^*[\phi](x) &:= Q^x(x)
}
\item[2.] For $t =1,2 \ldots T, i\in \cN, \forall a_{1:t-1}, x^i$
\seq{
\label{eq:mu*def}
\label{eq:beta*def}
\eq{
&\beta_t^{*,i}(a_t^i|a_{1:t-1},x^i)  := \theta_t^i[\pi_t^*[a_{1:t-1}]](a^i_t|x^i) \\
&^i\mu^{*}_{t}(x^{-i}|a_{1:t-1},x^i)   := \pi_t^*[a_{1:t-1}](x^{-i}|x^{i})\\
&\pi^*_{t+1}[a_{1:t}]  := F_t(\pi_t^{*}[a_{1:t-1}], \theta_t[\pi_t^*[a_{1:t-1}]],a_t)
}
}
\end{itemize}
where $F_t$ is defined in Lemma~\ref{lemma:C1}.

In the following theorem, we show that the equilibrium strategy and belief profile $(\beta^*,\mu^*)$ defined above constitute a PBE of the game considered.
\begin{theorem}
\label{Thm:Main}
A strategy and belief profile $(\beta^*,\mu^*)$, constructed through backward/forward recursive methodology is a PBE of the game, i.e.
$\forall i \in \cN,t \in \mathcal{T}, (a_{1:t-1},x^i), \beta_{t:T}^i$,
\eq{
&\E^{\beta_{t:T}^{*,i} \beta_{t:T}^{*,-i},\,\mu_t^{*}} \left\{ \sum_{n=t}^T R_n^i(X,A_n) \big\lvert  a_{1:t-1}, x^i \right\} \geq \nn\\
&\E^{\beta_{t:T}^{i} \beta_{t:T}^{*,-i},\, \mu_t^{*}} \left\{ \sum_{n=t}^T R_n^i(X,A_n) \big\lvert a_{1:t-1}, x^i \right\}. \label{eq:prop}
}
\end{theorem}
\begin{IEEEproof}
The proof is provided in Appendix~\ref{app:Theorem_Game}.
\end{IEEEproof}
\section{Remarks}
A few remarks are in order.

\begin{remark}
We note that in Step-2 in Backward Recursion, a fixed-point equation is solved in $(\tilde{\gamma}_t^i,V_t^i)_{i\in N}$ for each $\pi_t\in\Delta(\cX)$ and for each $t\in{T,T-1,\ldots1}$. Since Backward recursion dominates the computational complexity, the complexity of our methodology is linear in time. However, we do not make any claim about the complexity of solving the fixed-point in each instant, which, together with its existence as discussed below, is an important open question for future research.  
\end{remark}
\begin{remark}
We emphasize that even though the backward-forward methodology presented above finds a class of equilibrium strategies that are structured i.e. depend on the common belief and private state, the unilateral deviations of players in \eqref{eq:prop} are considered in the space of general strategies, i.e., the methodology does not make any bounded rationality assumptions.
\end{remark}

\begin{remark}
\underline{Intuition of the proof}: In such games, one could use the one-shot deviation principle~\cite{HeJaJoSl96} to argue that sequential rationality reduces to showing that no player wants to deviate unilaterally in $\beta_t^i$ at any time $t$, keeping the rest of the strategy $\beta^{i,*}_{t+1:T}$ as the equilibrium strategy. We argue that this is equivalent to~\eqref{eq:m_FP} i.e. for a given $(\pi_t^*,x^{i})$, a player's unilateral deviation in its strategy $\beta_t^{i,*}$ is the same as unilaterally deviation in its measure $\gamma_t^i(\cdot|x^i)$ on its action $a_t^i$ (and not on the whole function $\gamma_t^i(\cdot|\cdot)$). This is because under such unilateral deviations, a player uses the same future information states $(\pi_{t+1}^*,x^{i})$ as it would have done under an equilibrium strategy, whose update depends on equilibrium $\tilde{\gamma}_t$. This is so because player uses the same equilibrium $\pi_{t+1}^*$ to predict other player's actions, and update of its equilibrium private state $x^{i}$ does not depend on $\gamma_t^i$.
\end{remark}

\begin{remark}
\label{remark:existence}
While it is known that for any finite dynamic game with asymmetric information and perfect recall, there always exists a PBE~\cite[Prop. 249.1]{OsRu94}, existence of SPBE is not always guaranteed.
It is clear from our methodology that existence of SPBE boils down to existence of a solution to the fixed-point equation~\eqref{eq:m_FP} at every stage. Specifically, at each time $ t $ given the functions $ V_{t+1}^i $ for all $ i \in \cN $ from the previous round (in the backwards recursion) equation~\eqref{eq:m_FP} must have a solution $ \tilde{\gamma}_t^i $ for all $ i \in \cN $. Generally, existence of equilibria is shown through Kakutani's fixed point theorem, as is done by proving existence of a mixed strategy Nash equilibrium of a finite game~\cite{OsRu94,Na51}. This is done by showing existence of fixed point of the best-response correspondences of the game. Among other conditions, it requires the \emph{closed graph property} of the correspondences, which is usually implied by the continuity property of the utility functions involved.
For~\eqref{eq:m_FP} establishing existence is not straightforward due to: (a) potential discontinuity of the $\pi_t$ update function $F_t$ when the denominator in the Bayesian update is 0 and (b) potential discontinuity of the value functions, $V_{t+1}^i$. It is noted in~\cite[ch. 5]{Ta17} through~\cite{MiWe85} that for dynamic zero-sum games with asymmetric information, the value function is a continuous function, on the basis of which and the above mentioned arguments, the existence of equilibria in sequential decomposition fir such games is established. Furthermore, authors in~\cite{VaAn16cdc} consider an LQG dynamic game with asymmetric information (with linear state update, quadratic costs and Gaussian noise), where they present sufficient algorithmic conditions for such an equilibrium to exist. Authors in~\cite{VaSiAn17arxiv} study a public goods game for which signaling equilibria were found numerically using such methodology where agents have independent types which are perfectly observed by them and agents observe each others' actions in each time-period. Having said this, existence of the fixed-point equation for a more general class of problems remain an open question.
\end{remark}

\begin{remark}
This model also allows to incorporate many bounded rationality models. Some examples include using a discount factor $\delta$ or by restricting the search for optimum $\gamma_t^i(\cdot|x^i)$ in~\eqref{eq:m_FP} in the space of functions that are linear in private information variables. 
\end{remark}

\begin{remark}
In this paper, we considered a model with static types of the players. If the types of the players were also dynamic, say in a Markovian or a controlled Markovian way, we believe such a decomposition is not possible. The reason is that there is no consistent set of sufficient belief statistics that summarizes the observed history of the players in such a way that does not lead to \emph{infinite regress of beliefs}. We pose it as an open problem if there exists any special structure of dynamic evolution of private correlated states that allow for sequential decomposition in dynamic games with asymmetric information.
\end{remark}

\section{Conclusion}
\label{sec:Conclusion}
In this paper, we considered a model of repeated game where there is an underlying state of the system that is static and players jointly observe actions of other players with delay 1 and correlated private observations of the state of the system. Each player receives a reward that is a function of the state and actions of all the players. We define a common information belief state and private information belief states of the players. We then presented a backward-forward methodology similar to the one presented in~\cite{VaAn15arxiv} to compute its structured perfect Bayesian equilibria (SPBE). Future work includes proving such a methodology for discounted infinite-horizon case and specializing results to many practical settings such as games on graph, where players who are connected on the graph have correlated private information. Some practical applications of interest include security games for cyber-physical systems and Bayesian learning games in a social network with fully rational and potentially adversarial agents. An important future direction would to be investigate if this new result in the theory of such games can facilitate more efficient way of computing equilibria in~\cite{BrSa17} which can have significant implications in developing a software to solve real-world strategic problems.
 	\section{Acknowledgement}
The author would like to thank Achilleas Anastasopoulos for useful comments and pointing an error in previous draft, and Francois Baccelli and Sriram Vishwanath for the encouragement and support.
	\appendices
%
	\section{}
	\label{app:lemma_update}

\begin{lemma}
	There exists an update function $F_t$ of $\pi_t$, independent of $\theta$
	\eq{
	\pi_{t+1} = F_t(\pi_t,\gamma_t,a_{t}) \label{eq:piupdate}
	}
	\end{lemma}
	\begin{IEEEproof}
	\seq{
	\eq{
	&\pi_{t+1}(x) = P^{\theta}(x|a_{1:t},\gamma_{1:t+1}) \\
	&=  P^{\theta}(x|a_{1:t},\gamma_{1:t}) \\
	&= \frac{P^{\theta}(x,a_t|a_{1:t-1},\gamma_{1:t}) }
	{ \sum_{x,a_t}P^{\theta}(x,a_t|a_{1:t-1},\gamma_{1:t}) }\\
	&= \frac{ \pi_t(x)\left(\prod_{i=1}^N\gamma_t^i(a_t^i|x^i)\right)}{\sum_{x} \pi_t(x) \prod_{i=1}^N \gamma_t^i(a_t^i|x^i)}
	%
	%
	%
	}
	if the denominator is not 0 and 
	\eq{
	\pi_{t+1}(x)&= \pi_t(x)
	}
	otherwise. Thus we have,
\eq{
\pi_{t+1}	=  F_t(\pi_t, \gamma_t,a_t) ~\label{eq:F_update}
	}
	}
	\end{IEEEproof}
	
\section{(Proof of Theorem~\ref{Thm:Main})}
\label{app:Theorem_Game}
In the following theorem, we will assume that the equilibrium strategies and beliefs $(\beta^*,\mu^*)$ are generated using an equilibrium function $\theta$. Moreover, we will also use $\pi_t^*$ map corresponding to $^i\mu_t^*$ as defined in~\eqref{eq:mu*def}. With slight abuse of notation, we use both beliefs and belief functions as superscripts on expectations, where the reference is clear from the context. These functions when used as superscripts in expectation denote the belief functions, which when applied on the conditioned random variables, define beliefs on the random variables of interest inside the expectation.

\begin{IEEEproof}
We prove (\ref{eq:prop}) using induction and from results in Lemma~\ref{lemma:2}, \ref{lemma:3} and \ref{lemma:1} proved in Appendix~\ref{app:lemmas}. 
\seq{
For base case at $t=T$, $\forall i\in \cN, (a_{1:t-1},x^i)\in \mathcal{H}_{T}^i, \beta^i$
\eq{
&\E^{\beta_{T}^{*,i} \beta_{T}^{*,-i},\, \pi_{T}^{*} }\left\{  R_t^i(X,A_t) \big\lvert a_{1:t-1},x^i \right\}\nn\\
&=V^i_T(\pi_T^*[a_{1:t-1}], x^{i})  \label{eq:T2a}\\
&\geq \E^{\beta_{T}^{i} \beta_{T}^{*,-i},\, \pi_{T}^{*}} \left\{ R_t^i(X,A_t) \big\lvert a_{1:t-1},x^i \right\}  \label{eq:T2}
}
}
where (\ref{eq:T2a}) follows from Lemma~\ref{lemma:1} and (\ref{eq:T2}) follows from Lemma~\ref{lemma:2} in Appendix~\ref{app:lemmas}.

Let the induction hypothesis be that for $t+1$, $\forall i\in \cN, (a_{1:t}, x^i) \in \mathcal{H}_{t+1}^i, \beta^i$,
\seq{
\eq{
 & \E^{\beta_{t+1:T}^{*,i} \beta_{t+1:T}^{*,-i},\, \pi_{t+1}^{*}} \left\{ \sum_{n=t+1}^T R_n^i(X,A_n) \big\lvert  a_{1:t-1},x^i \right\} \geq\nn \\
 &\E^{\beta_{t+1:T}^{i} \beta_{t+1:T}^{*,-i},\, \pi_{t+1}^{*}} \left\{ \sum_{n=t+1}^T R_n^i(X,A_n) \big\lvert  a_{1:t-1},x^i \right\}. \label{eq:PropIndHyp}
}
}
\seq{
Then $\forall i\in \cN, (a_{1:t-1}, x^i)\in \mathcal{H}_t^i, \beta^i$, we have
\eq{
&\E^{\beta_{t:T}^{*,i} \beta_{t:T}^{*,-i},\, \pi_t^{*}} \left\{ \sum_{n=t}^T R_n^i(X,A_n) \big\lvert a_{1:t-1}, x^i \right\} \nonumber \\
&= V^i_t(\pi^*_t[a_{1:t-1}], x^{i})\label{eq:T1}\\
&\geq \E^{\beta_t^i \beta_t^{*,-i}, \,\pi_t^*} \left\{ R_t^i(X,A_t) + V_{t+1}^i (\pi^*_{t+1}[a_{1:t-1},A_{t}],x^{i}) \big\lvert a_{1:t-1}, x^i \right\}  \label{eq:T3}\\
&= \E^{\beta_t^i \beta_t^{*,-i}, \,\pi_t^*} \left\{ R_t^i(X,A_t) + \E^{\beta_{t+1:T}^{*,i} \beta_{t+1:T}^{*,-i},\, \pi_{t+1}^*}  \left\{ \sum_{n=t+1}^T R_n^i(X,A_n)  \big\lvert a_{1:t-1},A_{t}, x^i \right\}   \big\vert a_{1:t-1}, x^i \right\}  \label{eq:T3b}
}
\eq{
&\geq \E^{\beta_t^i \beta_t^{*,-i}, \,\pi_t^*} \left\{ R_t^i(X,A_t) + \E^{\beta_{t+1:T}^{i} \beta_{t+1:T}^{*,-i} \pi_{t+1}^{*}} \left\{ \sum_{n=t+1}^T R_n^i(X,A_n) \big\lvert a_{1:t-1},A_{t}, x^i\right\} \big\vert a_{1:t-1}, x^i \right\}  \label{eq:T4} 
}
\eq{
&= \E^{\beta_t^i \beta_t^{*,-i}, \,  \pi_t^*} \left\{ R_t^i(X,A_t)+ \E^{\beta_{t:T}^{i} \beta_{t:T}^{*,-i} \pi_t^{*}} \left\{ \sum_{n=t+1}^T R_n^i(X,A_n) \big\lvert a_{1:t-1},A_{t}, x^i\right\} \big\vert a_{1:t-1}, x^i \right\}  \label{eq:T5}\\
&=\E^{\beta_{t:T}^{i} \beta_{t:T}^{*,-i},\, \pi_t^{*}} \left\{ \sum_{n=t}^T R_n^i(X,A_n) \big\lvert a_{1:t-1},  x^i \right\}  \label{eq:T6},
}
}
where (\ref{eq:T1}) follows from Lemma~\ref{lemma:1}, (\ref{eq:T3}) follows from Lemma~\ref{lemma:2}, (\ref{eq:T3b}) follows from Lemma~\ref{lemma:1}, (\ref{eq:T4}) follows from induction hypothesis in (\ref{eq:PropIndHyp}) and (\ref{eq:T5}) follows from Lemma~\ref{lemma:3}.
\end{IEEEproof}

\section{}
\label{app:lemmas}
As we did in the previous theorem, in the following lemmas, we would assume that the equilibrium strategies and beliefs $(\beta^*,\mu^*)$ are generated using an equilibrium function $\theta$. We also use $\pi_t^*$ corresponding to $^i\mu_t^*$ as defined in~\eqref{eq:mu*def}.
\begin{lemma}
\label{lemma:2}
This lemma states that the reward that player $i$ would get on playing equilibrium strategy will be greater or equal to the reward it would get if it deviates only at time $t$, keeping the rest of its strategy as equilibrium strategy.
$\forall t\in \mathcal{T}, i\in \cN, (a_{1:t-1},x^i)\in \mathcal{H}_t^i, \beta^i_t$
\eq{
&V_t^i(\pi_t^*[a_{1:t-1}],x^i) \geq \E^{\beta_t^i\beta_t^{*,-i},\, \pi_t^*} \left\{ R_t^i(X,A_t) +  V_{t+1}^i (\pi_{t+1}^*[a_{1:t-1},A_{t}]) \big\lvert  a_{1:t-1},x^i \right\}\label{eq:lemma2}
}
\end{lemma}

\begin{IEEEproof}
We prove this lemma by contradiction.

Suppose the claim is not true for $t$. This implies $\exists i, \hat{\beta}_t^i, \ha_{1:t-1},x^i$ such that 
\eq{
&\E^{\hat{\beta}_t^i\beta_t^{*,-i},\, \pi_t^*} \left\{ R_t^i(X,A_t) + V_{t+1}^i (\pi_{t+1}^*[\ha_{1:t-1},A_{t}]) \big\lvert \ha_{1:t-1}, x^i \right\} \nn \\
&> V_t^i(\pi_t^*[\ha_{1:t-1}]).\label{eq:E8}
}
We will show that this contradicts the definition of  $V_t^i$ in (\ref{eq:Vdef}).\\
Construct $\hat{\gamma}^i_t(a_t^i|x^i) = \hat{\beta}_t^i(a_t^i|\ha_{1:t-1},x^i)  $

Then for $\ha_{1:t-1},x^i$, we have
\seq{
\eq{
&V_t^i(\pi_t^*[\ha_{1:t-1}]) \nn \\
&< \E^{\hat{\beta}_t^i,\beta_t^{*,-i}, \pi_t^*,{x}^{i}} \left\{ R_t^i(X,A_t) +  V_{t+1}^i (\pi_{t+1}^*[\ha_{1:t-1},A_t]) \big\lvert \ha_{1:t-1},x^i \right\}   \label{eq:E10}\\
&=\sum_{\substack{x,a_t,a_{t},\\x^i,x^{-i} }}\left[ R_t^i(X,A_t) + V_{t+1}^i (\pi_{t+1}[\ha_{1:t-1},a_{t}] ) \right] \pi_t^*[\ha_{1:t-1}](x^{-i}|{x}^{i})
 \hat{\beta}_t^i(a_t^i|\ha_{1:t-1},x^i) \beta_t^{*,-i}(a_t^{-i}|\pi_t^*[\ha_{1:t-1}] ,x^{-i}) \label{eq:E11}\\
&=\sum_{x,a_t,x^i,x^{-i}}\left[ R_t^i(X,A_t) +  V_{t+1}^i( \pi_{t+1}[\ha_{1:t-1},a_{t}] ) \right]\pi_t^*[\ha_{1:t-1}](x^{-i}|{x}^{i}) \hat{\gamma}_t^i(a_t^i|x^i) \beta_t^{*,-i}(a_t^{-i}|\pi_t^*[\ha_{1:t-1}] ,x^{-i})\label{eq:E11b}\\
&=\E^{\hat{\gamma}_t^i(\cdot|{x}^i) \beta_t^{*,-i},\,\pi_t^*[\ha_{1:t-1}]} \left\{ R_t^i(X,A_t) + V_{t+1}^i (\pi_{t+1}^*[\ha_{1:t-1},A_t]) \big\lvert {x}^{i}\right\} \\
&\leq \max_{\gamma^i_t(\cdot|{x}^i)} \E^{\gamma^i_t(\cdot|{x}^i) \beta_t^{*,-i}, \, \pi_t^*[\ha_{1:t-1}]} \left\{ R_t^i(X,A_t) +  V_{t+1}^i (\pi_{t+1}^*[\ha_{1:t-1},A_t]) \big\lvert {x}^{i}\right\}\\
&= V_t^i(\pi_t^*[\ha_{1:t-1}]) \label{eq:E12}
}
where (\ref{eq:E10}) follows from (\ref{eq:E8}), \eqref{eq:E11b} follows from the definition of $\hat{\gamma}_t^i$, and (\ref{eq:E12}) follows from the definition of $V_t^i$ in (\ref{eq:Vdef}).
}
\end{IEEEproof}

\begin{lemma}
\label{lemma:3}
$\forall i\in \cN, t\in \mathcal{T}, (a_{1:t},x^i)\in \mathcal{H}_{t+1}^i$ and
$\beta^i_t$
\eq{
&\E^{ \beta_{t:T}^{i}  \beta^{*,-i}_{t:T},\,\pi_t^{*}}  \left\{ \sum_{n=t+1}^T R_n^i(X,A_n) \big\lvert  a_{1:t}, x^i \right\} =\nn \\
&\E^{\beta^i_{t+1:T} \beta^{*,-i}_{t+1:T},\, \pi_{t+1}^{*}}  \left\{ \sum_{n=t+1}^T R_n^i(X,A_n) \big\lvert a_{1:t}, x^i \right\}. \label{eq:F1}
}
\end{lemma}
\begin{IEEEproof} 
Since the above expectations involve random variables $X^{-i}, A_{t+1:T}$, we consider\\
$P^{\beta^i_{t:T}\beta^{*,-i}_{t:T},\, \pi_t^{*}} (x ,a_{t+1:T} \big\lvert  a_{1:t},x^i )$.
\seq{
\eq{
&P^{\beta^i_{t:T} \beta^{*,-i}_{t:T},\, \pi_t^{*}} (x, a_{t+1:T}\big\lvert a_{1:t},x^i )
= P^{\beta^i_{t:T} \beta^{*,-i}_{t:T},\, \pi_t^{*}} (x\big\lvert a_{1:t},x^i )P^{\beta^i_{t:T} \beta^{*,-i}_{t:T},\, \pi_t^{*}} (a_{t+1:T}\big\lvert x,a_{1:t} ) \label{eq:F2}
}
We first note that
\eq{
&P^{\beta^i_{t:T} \beta^{*,-i}_{t:T},\, \pi_t^{*}} ( a_{t+1:T} \big\lvert  x, a_{1:t})\\
&= \beta_{t+1}^{i}(a_{t+1}^i| a_{1:t-1}, x^i) \left(\beta_{t+1}^{*,-i}(a_{t+1}^{-i}| a_{1:t-1},x^{-i} )\right)P^{\beta^i_{t:T} \beta^{*,-i}_{t:T},\, \pi_t^{*}} ( a_{t+2:T} \big\lvert a_{1:t},x )\\
&=P^{\beta^i_{t+1:T} \beta^{*,-i}_{t+1:T},\, \pi_{t+1}^{*}} (a_{t+1:T}| a_{1:t},x)
}
Thus using the above equation, (\ref{eq:F2}) is given by
\eq{
P^{\beta^i_{t+1:T} \beta^{*,-i}_{t+1:T},\, \pi_{t+1}^{*}} (x,a_{t+1:T}| a_{1:t-1},x^i)= P^{\beta_{t+1:T}^{ i} \beta_{t+1:T}^{*, -i},\, \pi_{t+1}^{*}}  (x,a_{t+1:T} |  a_{1:t-1},x^i ). \label{eq:1}
}
}

\end{IEEEproof}
\vspace{-0.3cm}
\begin{lemma}
\label{lemma:1}
$\forall i\in \cN, t\in \mathcal{T}, a_{1:t-1}\in \mathcal{H}_t^c, x^i\in (\cX^i)$

\eq{
&V^i_t(\pi^*_t[a_{1:t-1}],x^i)= \E^{\beta_{t:T}^{*,i} \beta_{t:T}^{*,-i},\pi_t^{*}} \left\{ \sum_{n=t}^T R_n^i(X,A_n) \big\lvert  a_{1:t-1}, x^i \right\} .
}
\end{lemma}

\begin{IEEEproof}

\seq{
We prove the lemma by induction. For $t=T$, 
\eq{
 &\E^{\beta_{T}^{*,i} \beta_{T}^{*,-i} ,\,\pi_{T}^{*}} \left\{  R_t^i(X,A_t) \big\lvert a_{1:t-1}, x^i \right\}\nn\\
 &=\sum_{x^{-i}, a_T,x^{-i}} R_t^i(X,A_t)\pi_{T}^{*}[a_{1:t-1}](x^{-i}|x^{i})\beta_{T}^{*,i}(a_T^i|a_{1:t-1},x^{i}) \beta_{T}^{*,-i}(a_T^{-i}|a_{1:t-1}, x^{-i})\\
 &=V^i_T(\pi^*_t[a_{1:t-1}], x^{i}) \label{eq:C1},
}
}
where (\ref{eq:C1}) follows from the definition of $V_t^i$ in (\ref{eq:Vdef}) and the definition of $\beta_T^*$ in the forward recursion in (\ref{eq:beta*def}).
Suppose the claim is true for $t+1$, i.e., $\forall i\in \cN, t\in \mathcal{T}, (a_{1:t-1},x^i)\in \mathcal{H}_{t+1}^i$
\eq{
&V^i_{t+1}(\pi^*_{t+1}[a_{1:t}], x^{i}) = \E^{\beta_{t+1:T}^{*,i} \beta_{t+1:T}^{*,-i},\, \pi_{t+1}^{*}} \left\{ \sum_{n=t+1}^T R_n^i(X,A_n) \big\lvert a_{1:t},x^i \right\} \label{eq:CIndHyp}.
}
Then $\forall i\in \cN, t\in \mathcal{T}, (a_{1:t-1},x^i)\in \mathcal{H}_t^i$, we have
	\seq{
\eq{
&\E^{\beta_{t:T}^{*,i} \beta_{t:T}^{*,-i} ,\,\pi_t^{*}} \left\{ \sum_{n=t}^T R_n^i(X,A_n) \big\lvert  a_{1:t-1}, x^i \right\} \nonumber \\
&=  \E^{\beta_{t:T}^{*,i} \beta_{t:T}^{*,-i} ,\,\pi_t^{*}} \left\{R_t^i(X,A_t) + \E^{\beta_{t:T}^{*,i} \beta_{t:T}^{*,-i} ,\,\pi_t^{*},x^{i}} \left\{ \sum_{n=t+1}^T R_n^i(X,A_n)\big\lvert a_{1:t-1}, A_t, x^i\right\} \big\lvert a_{1:t-1}, x^i \right\} \label{eq:C2}\\
&=  \E^{\beta_{t:T}^{*,i} \beta_{t:T}^{*,-i} ,\,\pi_t^{*}} \left\{R_t^i(X,A_t) +  \E^{\beta_{t+1:T}^{*,i} \beta_{t+1:T}^{*,-i},\, \pi_{t+1}^{*}} \left\{ \sum_{n=t+1}^T R_n^i(X,A_n)\big\lvert a_{1:t-1},A_{t}, x^i\right\} \big\lvert a_{1:t-1},x^i \right\} \label{eq:C3}\\
&=  \E^{\beta_{t:T}^{*,i} \beta_{t:T}^{*,-i} ,\,\pi_t^{*}} \left\{R_t^i(X,A_t) + V^i_{t+1}(\pi^*_{t+1}[a_{1:t-1},A_{t}], x^i) \big\lvert  a_{1:t-1},x^i \right\} \label{eq:C4}\\
&=  \E^{\beta_t^{*,i} \beta_t^{*,-i} ,\,\pi_t^{*}} \left\{R_t^i(X,A_t) +  V^i_{t+1}(\pi^*_{t+1}[a_{1:t-1},A_{t}], x^{i})\big\lvert  a_{1:t-1},x^i \right\} \label{eq:C5}\\
&=V^i_t(\pi^*_t[a_{1:t-1}],x^{i}) \label{eq:C6}
}
}
where (\ref{eq:C3}) follows from Lemma~\ref{lemma:3} in Appendix~\ref{app:lemmas}, (\ref{eq:C4}) follows from the induction hypothesis in (\ref{eq:CIndHyp}), (\ref{eq:C5}) follows because the random variables involved in expectation, $X,A_t$ do not depend on $\beta_{t+1:T}^{*,i} \beta_{t+1:T}^{*,-i}$ and (\ref{eq:C6}) follows from the definition of $V_t^i$ in (\ref{eq:Vdef}).
\end{IEEEproof}

\bibliographystyle{IEEEtran}

\begin{thebibliography}{10}
\providecommand{\url}[1]{#1}
\csname url@samestyle\endcsname
\providecommand{\newblock}{\relax}
\providecommand{\bibinfo}[2]{#2}
\providecommand{\BIBentrySTDinterwordspacing}{\spaceskip=0pt\relax}
\providecommand{\BIBentryALTinterwordstretchfactor}{4}
\providecommand{\BIBentryALTinterwordspacing}{\spaceskip=\fontdimen2\font plus
\BIBentryALTinterwordstretchfactor\fontdimen3\font minus
  \fontdimen4\font\relax}
\providecommand{\BIBforeignlanguage}[2]{{%
\expandafter\ifx\csname l@#1\endcsname\relax
\typeout{** WARNING: IEEEtran.bst: No hyphenation pattern has been}%
\typeout{** loaded for the language `#1'. Using the pattern for}%
\typeout{** the default language instead.}%
\else
\language=\csname l@#1\endcsname
\fi
#2}}
\providecommand{\BIBdecl}{\relax}
\BIBdecl

\bibitem{VaAn15arxiv}
\BIBentryALTinterwordspacing
D.~Vasal and A.~Anastasopoulos, ``A systematic process for evaluating
  structured perfect {B}ayesian equilibria in dynamic games with asymmetric
  information,'' Tech. Rep., Aug. 2015. [Online]. Available:
  \url{http://arxiv.org/abs/1508.06269}
\BIBentrySTDinterwordspacing

\bibitem{Ak70}
G.~A. Akerlof, ``The market for" lemons": Quality uncertainty and the market
  mechanism,'' \emph{The quarterly journal of economics}, pp. 488--500, 1970.

\bibitem{Sp73}
M.~Spence, ``Job market signaling,'' \emph{The quarterly journal of Economics},
  pp. 355--374, 1973.

\bibitem{FuTi91}
D.~Fudenberg and J.~Tirole, ``Perfect bayesian equilibrium and sequential
  equilibrium,'' \emph{journal of Economic Theory}, vol.~53, no.~2, pp.
  236--260, 1991.

\bibitem{Sietal17}
D.~Silver, J.~Schrittwieser, K.~Simonyan, I.~Antonoglou, A.~Huang, A.~Guez,
  T.~Hubert, L.~Baker, M.~Lai, A.~Bolton \emph{et~al.}, ``Mastering the game of
  go without human knowledge,'' \emph{Nature}, vol. 550, no. 7676, p. 354,
  2017.

\bibitem{BrSa17}
\BIBentryALTinterwordspacing
N.~Brown and T.~Sandholm, ``Safe and nested subgame solving for
  imperfect-information games,'' in \emph{Advances in Neural Information
  Processing Systems 30}, I.~Guyon, U.~V. Luxburg, S.~Bengio, H.~Wallach,
  R.~Fergus, S.~Vishwanathan, and R.~Garnett, Eds.\hskip 1em plus 0.5em minus
  0.4em\relax Curran Associates, Inc., 2017, pp. 689--699. [Online]. Available:
  \url{http://papers.nips.cc/paper/6671-safe-and-nested-subgame-solving-for-imperfect-information-games.pdf}
\BIBentrySTDinterwordspacing

\bibitem{Sh53}
L.~S. Shapley, ``Stochastic games,'' \emph{Proceedings of the national academy
  of sciences}, vol.~39, no.~10, pp. 1095--1100, 1953.

\bibitem{BaOl98}
T.~Ba{\c s}�ar and G.~Olsder, \emph{Dynamic Noncooperative Game Theory, 2nd
  Edition}.\hskip 1em plus 0.5em minus 0.4em\relax Society for Industrial and
  Applied Mathematics, 1998.

\bibitem{FiVr12}
J.~Filar and K.~Vrieze, \emph{Competitive Markov decision processes}.\hskip 1em
  plus 0.5em minus 0.4em\relax Springer Science \& Business Media, 2012.

\bibitem{OsRu94}
M.~J. Osborne and A.~Rubinstein, \emph{A Course in Game Theory}, ser. MIT Press
  Books.\hskip 1em plus 0.5em minus 0.4em\relax The MIT Press, 1994, vol.~1.

\bibitem{FuTi91book}
D.~Fudenberg and J.~Tirole, \emph{Game Theory}.\hskip 1em plus 0.5em minus
  0.4em\relax Cambridge, MA: MIT Press, 1991.

\bibitem{samuelson2006}
G.~J. Mailath and L.~Samuelson, \emph{Repeated games and reputations: long-run
  relationships}.\hskip 1em plus 0.5em minus 0.4em\relax Oxford university
  press, 2006.

\bibitem{MaTi01}
E.~Maskin and J.~Tirole, ``Markov perfect equilibrium: I. observable actions,''
  \emph{Journal of Economic Theory}, vol. 100, no.~2, pp. 191--219, 2001.

\bibitem{VaSiAn16arxiv}
D.~Vasal, A.~Sinha, and A.~Anastasopoulos, ``A systematic process for
  evaluating structured perfect bayesian equilibria in dynamic games with
  asymmetric information,'' \emph{IEEE Transactions on Automatic Control},
  2018.

\bibitem{VaAn16}
D.~Vasal and A.~Anastasopoulos, ``A systematic process for evaluating
  structured perfect {B}ayesian equilibria in dynamic games with asymmetric
  information,'' in \emph{American {C}ontrol {C}onference}, Boston, US, 2016,
  available on arXiv.

\bibitem{OuTaTe17}
Y.~Ouyang, H.~Tavafoghi, and D.~Teneketzis, ``Dynamic games with asymmetric
  information: Common information based perfect bayesian equilibria and
  sequential decomposition,'' \emph{IEEE Transactions on Automatic Control},
  vol.~62, no.~1, pp. 222--237, 2017.

\bibitem{Ta17}
H.~T. Jahormi, ``On design and analysis of cyber-physical systems with
  strategic agents,'' Ph.D. dissertation, University of Michigan, Ann Arbor,
  2017.

\bibitem{VaAn16cdc}
D.~Vasal and A.~Anastasopoulos, ``Signaling equilibria of dynamic {LQG} games
  with asymmetric information,'' in \emph{Conference on {D}ecision and
  {C}ontrol}, 2016.

\bibitem{VaAn16allerton}
\BIBentryALTinterwordspacing
------, ``Decentralized {B}ayesian learning in dynamic games,'' in
  \emph{Allerton Conference on Communication, Control, and Computing}, 2016.
  [Online]. Available: \url{https://arxiv.org/abs/1607.06847}
\BIBentrySTDinterwordspacing

\bibitem{TaOuTe16cdc}
H.~Tavafoghi, Y.~Ouyang, and D.~Teneketzis, ``On stochastic dynamic games with
  delayed sharing information structure,'' in \emph{Decision and Control (CDC),
  2016 IEEE 55th Conference on}.\hskip 1em plus 0.5em minus 0.4em\relax IEEE,
  2016, pp. 7002--7009.

\bibitem{KuVa86}
P.~Kumar and P.~Varaiya, ``Stochastic systems,'' 1986.

\bibitem{Ho80}
Y.-C. Ho, ``Team decision theory and information structures,''
  \emph{Proceedings of the IEEE}, vol.~68, no.~6, pp. 644--654, 1980.

\bibitem{NaMaTe13}
A.~Nayyar, A.~Mahajan, and D.~Teneketzis, ``Decentralized stochastic control
  with partial history sharing: A common information approach,''
  \emph{Automatic Control, IEEE Transactions on}, vol.~58, no.~7, pp.
  1644--1658, 2013.

\bibitem{Ma13}
A.~Mahajan, ``Optimal decentralized control of coupled subsystems with control
  sharing,'' \emph{Automatic Control, IEEE Transactions on}, vol.~58, no.~9,
  pp. 2377--2382, 2013.

\bibitem{MaTe08}
A.~Mahajan and D.~Teneketzis, ``On the design of globally optimal communication
  strategies for real-time communcation systems with noisy feedback,''
  \emph{IEEE J.~Select.~Areas Commun.}, no.~4, pp. 580--595, May 2008.

\bibitem{VaAn14}
D.~Vasal and A.~Anastasopoulos, ``Stochastic control of relay channels with
  cooperative and strategic users,'' \emph{IEEE Transactions on
  Communications}, vol.~62, no.~10, pp. 3434--3446, Oct 2014.

\bibitem{HeJaJoSl96}
E.~Hendon, H.~J. Jacobsen, and B.~Sloth, ``The one-shot-deviation principle for
  sequential rationality,'' \emph{Games and Economic Behavior}, vol.~12, no.~2,
  pp. 274--282, 1996.

\bibitem{Na51}
J.~Nash, ``Non-cooperative games,'' \emph{Annals of mathematics}, pp. 286--295,
  1951.

\bibitem{MiWe85}
P.~R. Milgrom and R.~J. Weber, ``Distributional strategies for games with
  incomplete information,'' \emph{Mathematics of operations research}, vol.~10,
  no.~4, pp. 619--632, 1985.

\bibitem{VaSiAn17arxiv}
\BIBentryALTinterwordspacing
D.~Vasal, A.~Sinha, and A.~Anastasopoulos, ``A systematic process for
  evaluating structured perfect {B}ayesian equilibria in dynamic games with
  asymmetric information,'' Tech. Rep., Aug. 2015. [Online]. Available:
  \url{http://arxiv.org/abs/1508.06269v3}
\BIBentrySTDinterwordspacing

\end{thebibliography}


\begin{thebibliography}{}
\providecommand{\url}[1]{#1}
\csname url@samestyle\endcsname
\providecommand{\newblock}{\relax}
\providecommand{\bibinfo}[2]{#2}
\providecommand{\BIBentrySTDinterwordspacing}{\spaceskip=0pt\relax}
\providecommand{\BIBentryALTinterwordstretchfactor}{4}
\providecommand{\BIBentryALTinterwordspacing}{\spaceskip=\fontdimen2\font plus
\BIBentryALTinterwordstretchfactor\fontdimen3\font minus
  \fontdimen4\font\relax}
\providecommand{\BIBforeignlanguage}[2]{{%
\expandafter\ifx\csname l@#1\endcsname\relax
\typeout{** WARNING: IEEEtran.bst: No hyphenation pattern has been}%
\typeout{** loaded for the language `#1'. Using the pattern for}%
\typeout{** the default language instead.}%
\else
\language=\csname l@#1\endcsname
\fi
#2}}
\providecommand{\BIBdecl}{\relax}
\BIBdecl

\end{thebibliography}

\end{document}